\documentclass[doublecol]{epl2} 

\usepackage{graphicx}
\usepackage{dcolumn}
\usepackage{bm}
\usepackage{hyperref}
\usepackage{color}
\usepackage{lipsum}
\usepackage{SIunits}
\usepackage{cite}

\title{Drop trampoline}

\author{Pierre Chantelot \and Martin Coux \and Christophe Clanet \and David Quere}
\shortauthor{Pierre Chantelot \etal}

\institute{                    
  \inst{1} PMMH, UMR 7636 du CNRS, ESPCI Paris, 75005 Paris, France\\
  \inst{2} LadHyX, UMR 7646 du CNRS, Ecole polytechnique, 91128 Palaiseau, France
}
\pacs{nn.mm.xx}{First pacs description}
\pacs{nn.mm.xx}{Second pacs description}
\pacs{nn.mm.xx}{Third pacs description}

\abstract{
Rigid superhydrophobic materials have the ability to repel millimetric water drops, in typically 10 ms. Yet, most natural water-repellent materials can be deformed by impacting drops. 
To test the effect of deformability, we perform impacts of non-wetting drops onto thin ($\sim$ 10 \micro m), circular PDMS membranes.
The bouncing mechanism is markedly modified compared to that on a rigid material: the liquid leaves the substrate as it is kicked upwards by the membrane.
We show that the rebound is controlled by an interplay between the dynamics of the drop and that of the soft substrate, so that we can continuously vary the contact time by playing on the membrane's characteristics and reduce it up to 70\%.
}

\begin{document}

\maketitle

Liquid droplets and soft solids can interact in many interesting ways. For instance, elastocapillary effects can lead to deformation of solids close to contact lines \cite{carre1996,Marchand2012} or to bending of slender and thin structures \cite{bico2004,py2007}. In more dynamical situations, substrate deformation can induce self-propulsion \cite{style2013,karpitschka2016}, prevent splashing \cite{howland2016,pepper2008} or allow liquid penetration in soft solids \cite{tagawa2013}.
It can also lead to improvement of the water or ice repellency of superhydrophobic materials \cite{Vasileiou2016, vasileiou2017}. 
Such materials have the ability to reflect impacting drops and we focus here on the way their flexibility affects this property. The contact time  $\tau$ of millimetric water drops on a rigid, repellent solid is on the order of 10 ms, which can be large enough to induce freezing \cite{mishchenko2010}, significant heat transfer \cite{Shiri2017} or contamination by surfactants \cite{song2017}. 
Different techniques have been proposed to reduce $\tau$. Decorating the substrate with macrotextures (such as ridges) was found to divide $\tau$ by a factor of typically 2 \cite{Bird2013}, a reduction also obserbved using soft membranes, as recently shown by Weisensee \emph{et al.} \cite{Weisensee2016}. This effect occurs at large impact velocity, in a regime difficult to explore due to splashing, which might explain the scattered nature of the results. Vasileiou \emph{et al.} also stressed the ability of soft membranes to reflect viscous drops - a point of obvious practical interest - but did not provide neither a specific study on the contact time nor a model to account for its reduction \cite{vasileiou2017}.
In order to study systematically the ability of soft solids to enhance water repellency, we chose to texture liquids rather than solids, that is, to use liquid marbles as a model of non-wetting drops \cite{aussillous2006}.
This allows us to show that the interplay between flexible substrates and non-wetting impacts leads to the possibility of  continuously tuning the contact time, which we model.
\begin{figure*}
\includegraphics[width=\textwidth]{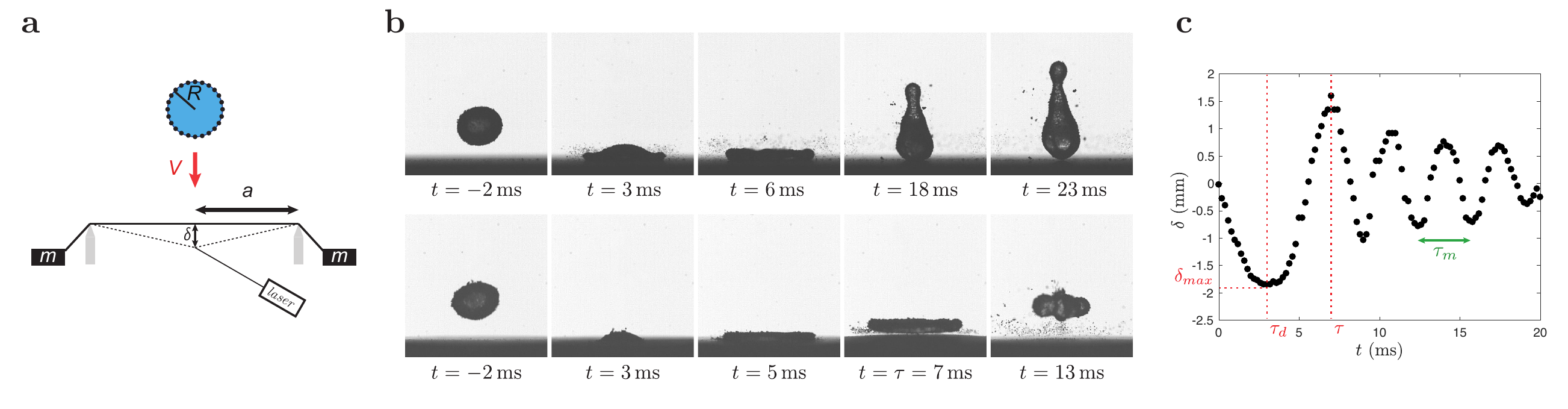}
\caption{\label{fig1}\textbf{a} A liquid marble (with radius $R$ and velocity $V$) impacts a circular PDMS membrane with radius $a$ and thickness $20$ \micro m put into tension by a mass $m$. The membrane is superhydrophobic and its deflection $\delta$ is measured by the deviation of an oblique laser sheet. \textbf{b} (top) A liquid marble ($R=1.8$ mm) impacts a rigid superhydrophobic substrate at $V=0.75$ m/s. The drop leaves the substrate after 22.5 ms. (bottom) Same liquid marble impacting a flexible membrane ($a = 10$ mm, $m = 12.23$ g) at $V=1$ m/s. The drop is kicked off after 7 ms, a reduction of 70 \% compared to the rigid case, with a pancake shape. \textbf{c} Deflection $\delta$ of the center of the membrane for the impact shown in b, from which we obtain the maximal deflection $\delta_{max}$ and its time $\tau_d$. Later, the free oscillations of the membrane give access to its fundamental period $\tau_m$.}
\end{figure*}

The experiment is sketched in figure \ref{fig1}a.
Our substrates are polydimethylsiloxane (PDMS) sheets with thickness $h=20$ \micro m (Silex) clamped between two plexiglas rings. The clamped sheets are placed on a frame with radius $a$ ($a = 7.5, 10, 17.5, 25$ mm) and tension is adjusted by weighting the membrane with a mass $m$. 
Liquid marbles are made by coating distilled water (density $\rho = 1000$ kg/m$^3$ and surface tension $\gamma=72$ mN/m) with lycopodium grains (diameter $\sim30$ \micro m) treated with fluoro-decyl-trichlorosilane. These marbles have a surface tension $\gamma \simeq 57 \pm 8$ mN/m slightly smaller than that of water \cite{planchette2012,aussillous2006}, and non-wet any substrate on which they are placed. They can resist impacts of moderate velocity, a property enhanced on a repellent substrate. To that end, the membranes are made water-repellent by spraying a solution of hydrophobic nanobeads dispersed in acetone (Ultra Ever Dry, UltraTech International). After evaporation of the solvent, the PDMS is coated by a few layers of nanobeads and it exhibits the high contact angles ($\sim160^{\circ}$) and low hysteresis ($\sim6^{\circ}$) typical from superhydrophobic materials.
Water drops ($R = 1$ mm and $R = 1.8$ mm) made from calibrated needles roll on a horizontal groove covered by lycopodium grains, until they get coated. Then we release them above the substrate through a hole at the end of the groove. The impact velocity $V$ can be varied between 0.5 and 1.5 m/s by adjusting the height from which these marbles fall. 
 We record side views of the impact and monitor the membrane deflection $\delta$ through the observation of a laser sheet in oblique incidence using two fast video cameras (Phantom V7) working at typically 10000 frames per second. The vertical position of the membrane is directly proportionnal to the displacement of the laser sheet. 

Marbles bounce off flexible substrates differently from what they do on rigid materials (figure \ref{fig1}b). In the latter case (top sequence), they spread, recoil and take off with an elongated shape, here after 22.5 ms. The contact time $\tau$ of a marble with radius $R=1.8$ mm is roughly independant of the impact velocity $V$ and plateaus at $\tau_0 = 22.7\pm0.8$ ms (see supplementary figure SI1), a value comparable to that of a non-wetting drop. 
In contrast (bottom sequence), a marble impacting a flexible membrane deforms its substrate, which renders it invisible in our side views. As the membrane recovers its horizontality, we observe that the spread marble takes off with a flattened shape, which suggests a reduction of contact time. $\tau$ in figure \ref{fig1}b is $7$ ms, that is, about one third of $\tau_0$, a reduction even larger than that reported in \cite{Weisensee2016, vasileiou2017}. Recoiling takes place later, while the drop is in the air.
Figure \ref{fig1}c shows the time evolution of the deflection $\delta$ of the center of the membrane. Firstly, the membrane sinks down to its maximal deflection $\delta_{max}$ at time $\tau_d$ (see supplementary figure SI2). Then, the substrate moves back and goes above the horizontal, which kicks the marble and makes it take off (at time $\tau$) at the membrane's uppermost position.
Later, it freely oscillates, allowing us to measure its natural period, here $\tau_m=3.45 \pm 0.10$ ms. These free oscillations are faster than the first oscillation forced by impact, showing that our system has a characteristic time $\tau$ intermediate between $\tau_m$ and $\tau_0$, the respective response times of the membrane and of the drop.\\ 

The time $\tau_m$ (and corresponding frequency $f_m=1/\tau_m$) can be varied by tuning the membrane geometry (through the radius $a$) and tension (through the mass $m$, see supplementary figure SI3). We show in figure \ref{fig3}a how the contact time $\tau$ varies as a function of the impact velocity $V$ for various frequencies $f_m$. For each value of $f_m$, the contact time is roughly independant of the impact velocity $V$, apart from a weak increase at low $V$ also observed for drops on rigid substrates \cite{Chevy2012}.
More importantly, we confirm our main observation: the contact time $\tau$ on soft membranes is reduced compared to $\tau_0$, the plateau value on a rigid substrate indicated in figure \ref{fig3}a with a dashed line. Specifically, $\tau$ decreases as we increase the frequency of the membrane, showing the influence of the response of the substrate on the timescale at which the liquid is repelled. 
The marble size $R$ also influences the contact time, small drops being shed faster than large ones, as shown in supplementary figure SI4 where it is observed that $\tau$ varies slower than $R^{3/2}$, the usual inertio-capillary behavior \cite{richard2002}. 
The response of the substrate can be characterized spatially, and we plot in figure \ref{fig3}b the maximal deflection $\delta_{max}$ as a function of $V$. $\delta_{max}$ varies linearly with $V$, and its value is typically millimetric.
When $R$ is fixed (filled circles, $R=1.8$ mm), there is no obvious relationship between $\delta_{max}$ and $f_m$. At fixed $f_m$ ($f_m=290$ Hz, green circles and triangles), $\delta_{max}$ increases with $R$, a logical consequence of the change in liquid mass. \\

Our aim is to understand how the liquid and the membrane cooperate in an original bouncing mechanism. Our analysis holds for $\tau_m<\tau_0$, the only regime where we expect contact time reduction. We model the solid/liquid system as coupled oscillators (figure \ref{fig4}a and \ref{fig4}b).
\begin{figure}
\centering
\includegraphics[width=0.4\textwidth]{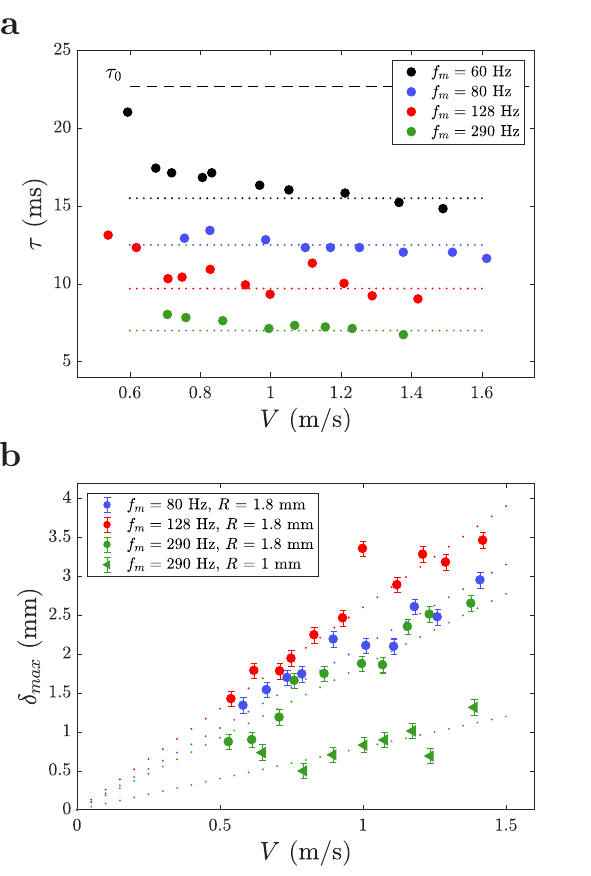}
\caption{\label{fig3}\textbf{a} Contact time $\tau$ of liquid marbles with radius $R=1.8$ mm bouncing off flexible membranes with frequencies $f_m$ as a function of impact velocity $V$. The dashed line represents the contact time $\tau_0$ on a rigid substrate. \textbf{b} Maximal deflection $\delta_{max}$ at the center of the membrane as a function of $V$. The dotted lines are linear fits.}
\end{figure}
On the one hand, the marble can be represented as a spring of stiffness $k_d$ and a mass $m_d$, a system with an oscillating frequency $f_d = \frac{1}{2\pi}\sqrt{\frac{k_d}{m_d}}$. Lord Rayleigh \cite{rayleigh1879} calculated the frequency of freely oscillating drops, and showed that it writes: $f_d = \sqrt{\frac{8\gamma}{3\pi m_d}}$, which provides the stiffness $k_d$ of the spring: $k_d = \frac{32\pi}{3}\gamma$. On the other hand, the membrane can be modelled as a spring of stiffness $k$, mass $m_m$ and fundamental frequency $f _m = \frac{1}{2\pi}\sqrt{\frac{k}{m_m}}$. 
We assume that the droplet-membrane system behaves during contact as oscillators in series, as sketched in figure \ref{fig4}b. Then the position $z$ of the membrane obeys a 4th-order differential equation:
\begin{equation}
\frac{d^4z}{dt^4} + \frac{k_d}{m_m}(1+\frac{m_m}{m_d}+\frac{k}{k_d})\frac{d^2z}{dt^2} + \frac{kk_d}{m_dm_m}z =0
\label{eq1}
\end{equation}
Equation \ref{eq1} has two natural limits. (1) On a rigid substrate ($k \rightarrow \infty$), it reduces to: $\ddot{z} + 4\pi^2f_d^2z=0$. The contact time on rigid repellent materials is simply proportional to the Rayleigh period $1/f_d$ \cite{richard2002}.
(2) A rigid bead ($k_d \rightarrow \infty$) hitting a flexible membrane is described by the equation $\ddot{z}+4\pi^2f_b^2z = 0$, with $f_b = \frac{1}{2\pi}\sqrt{\frac{k}{m_d+m_m}}$, that is, the frequency of a membrane of stiffness $k$ and mass $m_d+m_m$. We performed experiments with polypropylene beads ($R_b = 1$ mm and $R_b=1.75$ mm , $\rho_b = 900$ kg/m$^3$) with mass $m_b$, and our data plotted in figure \ref{fig4}c show that the reduced contact time $\tau f_b$ collapses on a single curve, confirming that $f_b$ is the frequency of the bead-membrane system. However this added mass argument does not capture the contact time reduction observed for drops as shown in the supplementary figure SI5. \\ 

Coming back to the impact of a water drop on a flexible substrate, we can notice that equation \ref{eq1} provides two natural frequencies, that is, $f_* = \frac{1}{2\pi}(\frac{k_d}{m_m}(1+\frac{m_m}{m_d}+\frac{k}{k_d}))^{1/2}$ and $f = \frac{1}{2\pi}(\frac{kk_d}{m_dm_m})^{1/4}$.
As shown in the SI, we have $f_* > f$ whatever the values of the physical parameters, which suggests that the dynamics of the system is set by $1/f$, the longer timescale.
When we rescale the contact time $\tau$ by the frequency $f$ and plot it as a function of the impact velocity $V$ (figure \ref{fig4}d), data for various $f_m$ (such as in figure \ref{fig3}a) and various $R$ indeed collapse. Apart from the increase observed at low $V$ \cite{Chevy2012}, contact time is found to plateau at a value $\tau \sim 0.75 f$. The frequency $f$ turns out to be the geometric mean of that of the drop and of the membrane, $f = \sqrt{f_m f_d}$, a formula capturing how the two objects conspire to generate fast bouncing. Interestingly, the frequency $f$ scales as $R^{-3/4}$, a behavior very different from that on a rigid substrate (where it varies as $R^{-3/2}$), in agreement with our results in the supplementary figure SI4.
Knowing the frequency $f$ yields a simple prediction for $\delta_{max}$.
Before impact, the membrane is immobile and the droplet with mass $m_d$ moves at speed $V$; during the first oscillation, drop and membrane both oscillate at the frequency $f$. Conserving the momentum provides the following scaling: $m_d V \sim (m_m + m_d)\delta_{max} f$. Figure \ref{fig4}e and \ref{fig4}f represent $\delta_{max} f_b (1+\frac{m_m}{m_b})$ and $\delta_{max} f (1+\frac{m_m}{m_d})$ as a function of $V$. For both solid and liquid marbles, we observe the predicted linear relationship, with respective slopes 0.4 $\pm$ 0.1 and 0.33 $\pm$ 0.03. These slopes are smaller than 1, suggesting that part of the initial momentum is not injected into membrane oscillations but also in membrane stretching and dissipation in air, and for liquids in internal motion.\\
\begin{figure*}[t]
\centering
\includegraphics[height=0.42\textheight]{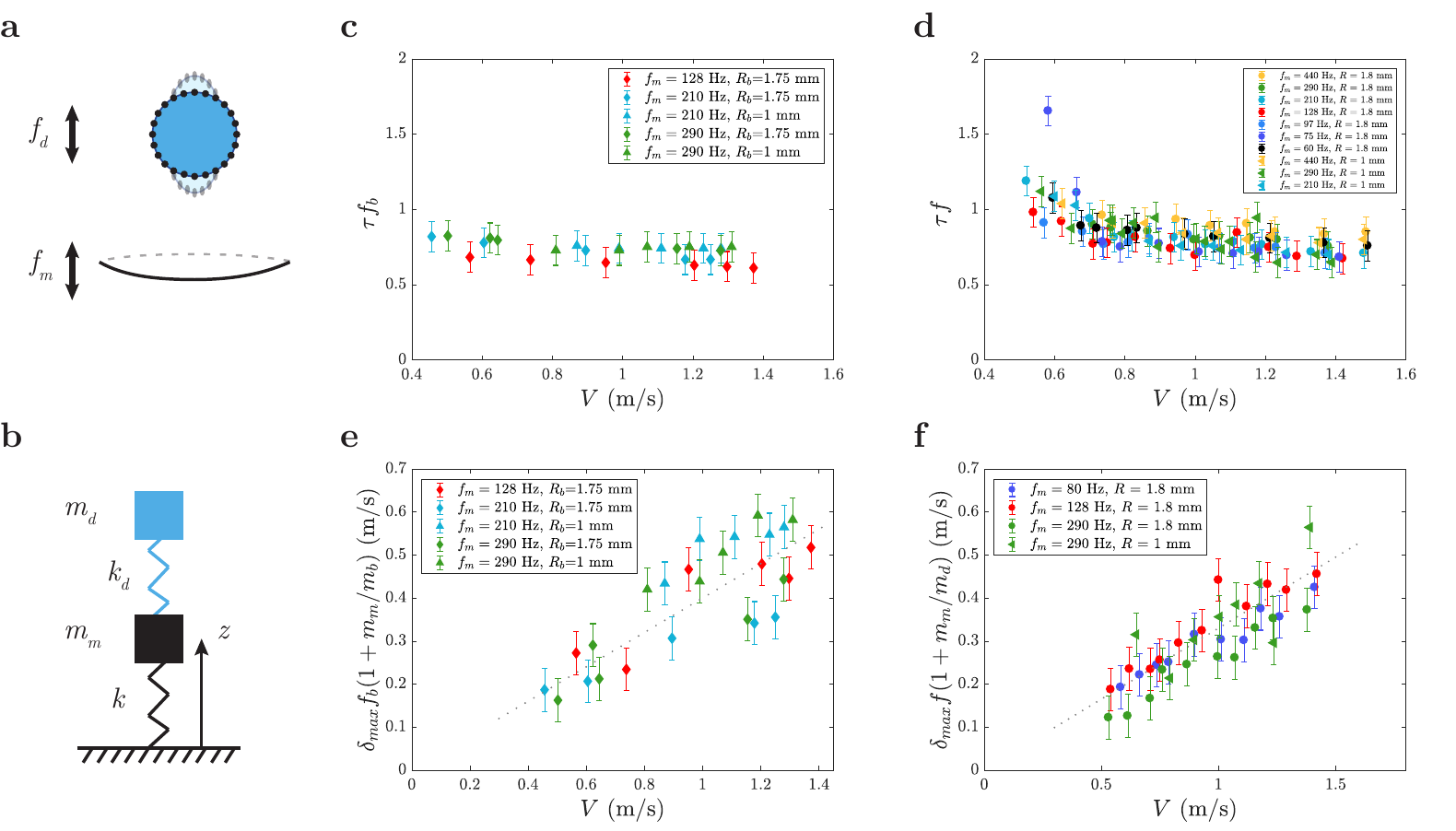}
\caption{\label{fig4}\textbf{a} The drop and membrane are modelled as oscillators of respective frequencies $f_d$ and $f_m$. \textbf{b} We assume that this system behaves as two oscillators in series during contact. \textbf{c} Normalized contact time $\tau f_b$ for solid beads impacting a flexible membrane at velocity $V$. $f_b = \sqrt{\frac{k}{m_m + m_d}}$ is the frequency of a membrane with mas $m_d + m_m$. \textbf{d} Normalized contact time $\tau f$ of liquid marbles impacting a flexible membrane at velocity $V$ where $f=\sqrt{f_d f_m}$ is extracted from equation \ref{eq1}. \textbf{e} Maximum deflection $\delta_{max}$ of the membrane as a function of the impact velocity $V$ for various membrane frequencies and bead radii. Collapse of data is obtained by multiplying $\delta_{max}$ by the quantity $f_{b} (1+\frac{m_m}{m_b})$  as suggested in the text. \textbf{f} Same plot for marble impact. $\delta_{max}$ is now multiplied by $f (1+\frac{m_m}{m_d})$. In both cases we observe a linear behavior as predicted in the text. Dotted lines are linear fits with slopes 0.4 $\pm$ 0.1 and 0.33 $\pm$ 0.03, respectively.}
\end{figure*}

Making impacting drops interact with soft solids modifies the outcome of collisions and can lead to enhanced repellency expressed by a systematic reduction of contact time.
The timescale of such impacts depends on the dynamics of both membrane and drop, so that we can continuously adjust the contact time between $0.3\tau_0$ (a very low value) and $\tau_0$, by playing on the membrane characteristics and the drop radius.
The opposite case ($f_m < f_d$) would deserve a separate study. On the one hand, we classically expect in this limit a Rayleigh bouncing time scaling as $1/f_d$. On the other hand, the soft nature of the substrate implies that the membrane should be deformed at impact and takeoff, making it usable as a force sensor for drop impacts. 

\acknowledgments
We thank Direction G\'{e}n\'{e}rale de l'Armement (DGA) for contributing to the financial support. We also thank Hadrien Bense, Lucie Domino and Beno\^{i}t Roman for insightful comments.

\bibliography{biblio_trampo}

\providecommand{\noopsort}[1]{}\providecommand{\singleletter}[1]{#1}%
\begin{thebibliography}{10}
\expandafter\ifx\csname url\endcsname\relax\def\url#1{\texttt{#1}}\fi

\bibitem{carre1996}
\Name{Carr{\'e} A., Gastel J.-C. \and Shanahan M.~E.}
  \REVIEW{Nature}{379}{1996}{432}.

\bibitem{Marchand2012}
\Name{Marchand A., Das S., Snoeijer J.~H. \and Andreotti B.} \REVIEW{Physical
  Review Letters}{109}{2012}{236101}.

\bibitem{bico2004}
\Name{Bico J., Roman B., Moulin L. \and Boudaoud A.}
  \REVIEW{Nature}{432}{2004}{690}.

\bibitem{py2007}
\Name{Py C., Reverdy P., Doppler L., Bico J., Roman B. \and Baroud C.~N.}
  \REVIEW{Physical Review Letters}{98}{2007}{156103}.

\bibitem{style2013}
\Name{Style R.~W., Che Y., Park S.~J., Weon B.~M., Je J.~H., Hyland C., German
  G.~K., Power M.~P., Wilen L.~A., Wettlaufer J.~S. \etal} \REVIEW{Proceedings
  of the National Academy of Sciences}{110}{2013}{12541}.

\bibitem{karpitschka2016}
\Name{Karpitschka S., Pandey A., Lubbers L.~A., Weijs J.~H., Botto L., Das S.,
  Andreotti B. \and Snoeijer J.~H.} \REVIEW{Proceedings of the National Academy
  of Sciences}{113}{2016}{7403}.

\bibitem{howland2016}
\Name{Howland C.~J., Antkowiak A., Castrej{\'o}n-Pita J.~R., Howison S.~D.,
  Oliver J.~M., Style R.~W. \and Castrej{\'o}n-Pita A.~A.} \REVIEW{Physical
  Review Letters}{117}{2016}{184502}.

\bibitem{pepper2008}
\Name{Pepper R.~E., Courbin L. \and Stone H.~A.} \REVIEW{Physics of
  Fluids}{20}{2008}{082103}.

\bibitem{tagawa2013}
\Name{Tagawa Y., Oudalov N., El~Ghalbzouri A., Sun C. \and Lohse D.}
  \REVIEW{Lab on a Chip}{13}{2013}{1357}.

\bibitem{Vasileiou2016}
\Name{Vasileiou T., Gerber J., Prautzsch J., Schutzius T.~M. \and Poulikakos
  D.} \REVIEW{Proceedings of the National Academy of
  Sciences}{113}{2016}{13307}.

\bibitem{vasileiou2017}
\Name{Vasileiou T., Schutzius T.~M. \and Poulikakos D.}
  \REVIEW{Langmuir}{33}{2017}{6708}.

\bibitem{mishchenko2010}
\Name{Mishchenko L., Hatton B., Bahadur V., Taylor J.~A., Krupenkin T. \and
  Aizenberg J.} \REVIEW{ACS Nano}{4}{2010}{7699}.

\bibitem{Shiri2017}
\Name{Shiri S. \and Bird J.~C.} \REVIEW{Proceedings of the National Academy of
  Sciences}{114}{2017}{6930}.

\bibitem{song2017}
\Name{Song M., Ju J., Luo S., Han Y., Dong Z., Wang Y., Gu Z., Zhang L., Hao R.
  \and Jiang L.} \REVIEW{Science Advances}{3}{2017}{}.

\bibitem{Bird2013}
\Name{Bird J.~C., Dhiman R., Kwon H.-M. \and Varanasi K.~K.}
  \REVIEW{Nature}{503}{2013}{385}.

\bibitem{Weisensee2016}
\Name{Weisensee P.~B., Tian J., Miljkovic N. \and King W.~P.}
  \REVIEW{Scientific Reports}{6}{2016}{1}.

\bibitem{aussillous2006}
\Name{Aussillous P. \and Qu{\'e}r{\'e} D.} \REVIEW{}{462}{2006}{}.

\bibitem{planchette2012}
\Name{Planchette C., Lorenceau E. \and Biance A.-L.} \REVIEW{Soft
  Matter}{8}{2012}{2444}.

\bibitem{Chevy2012}
\Name{Chevy F., Chepelianskii A., Qu{\'{e}}r{\'{e}} D. \and Rapha{\"{e}}l E.}
  \REVIEW{Europhysics Letters}{100}{2012}{54002}.

\bibitem{richard2002}
\Name{Richard D., Clanet C. \and Qu{\'{e}}r{\'{e}} D.}
  \REVIEW{Nature}{417}{2002}{811}.

\bibitem{rayleigh1879}
\Name{Rayleigh L.} \REVIEW{}{29}{1879}{}.

\end{thebibliography}
\bibliographystyle{eplbib}
\end{document}


\maketitle

\section{Contact time of liquid marbles}
\begin{figure}[h]
\centering
\includegraphics[width = 0.4\textwidth]{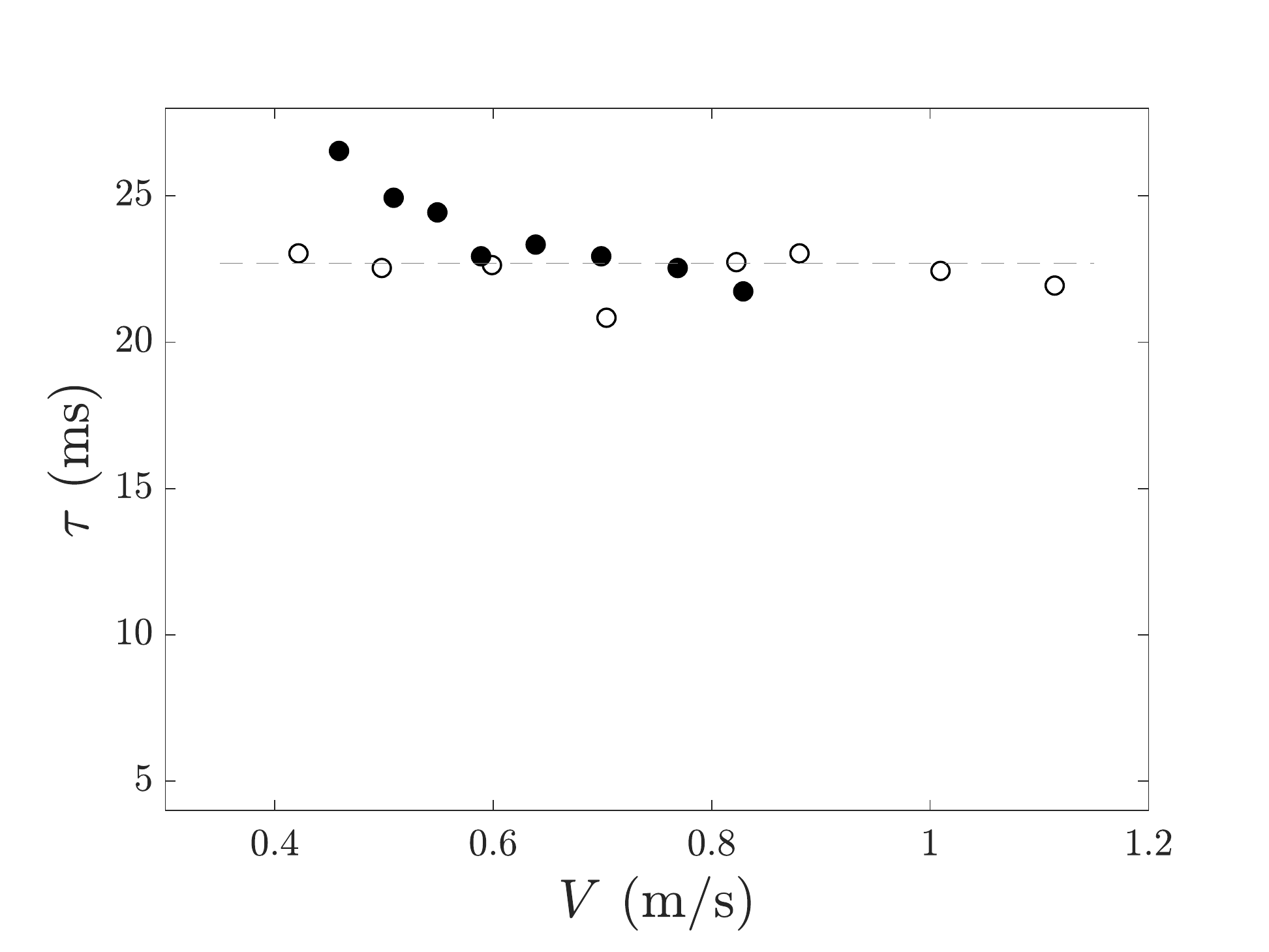}
\caption{\label{fig2_SI}Contact time $\tau$ of liquid marbles (black dots) and water droplets (black circles) with radius $R=1.8$ mm as a function of impact velocity $V$. Above 0.5 m/s for marbles, $\tau$ is independant of $V$ and we measure $\tau_0=22.7$ $\pm$ 0.8 ms.}
\end{figure}

We measure the contact time of liquid marbles impacting a rigid superhydrophobic substrate (made water repellent by spray coating of Ultra Ever Dry) as a function of the impact velocity $V$. Figure \ref{fig2_SI} shows the contact time $\tau$ of water droplets and marbles (black circles and dots respectively) as a function of $V$. Above 0.5 m/s, $\tau$ does not depend on $V$ for marbles and droplets and takes roughly the same values. We measure the plateau value $\tau_0 = 22.7 \pm 0.8$ ms for marbles. Below 0.5 m/s, the contact time of marbles increases in a similar fashion to what is observed for droplets below 0.3 m/s.

\section{Maximal deflection time $\tau_d$}
\begin{figure}[h]
\centering
\includegraphics[width = 0.9\textwidth]{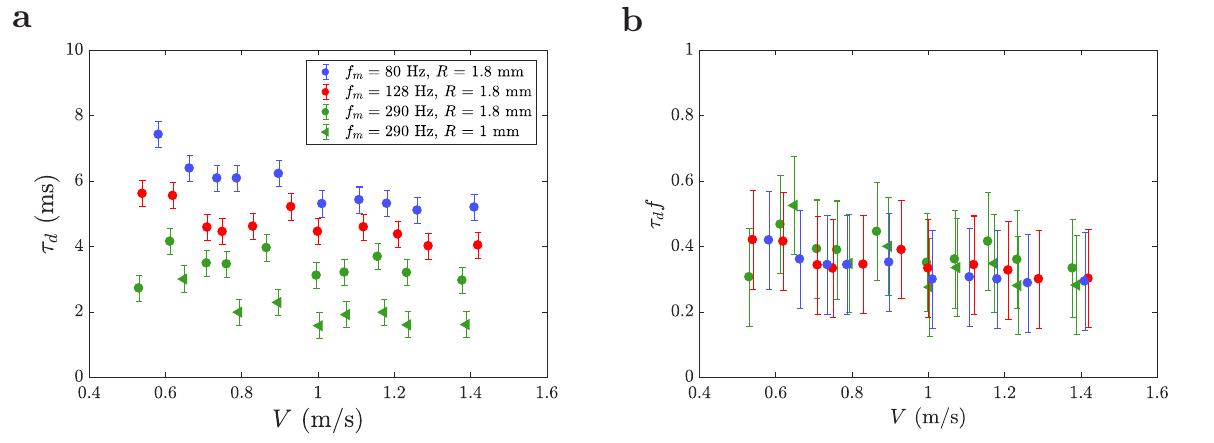}
\caption{\label{fig3_SI}\textbf{a} Deflection time $\tau_d$ as a function of impact velocity $V$ for membranes with different frequencies $f_m$ (dots) and marbles with different radius $R$. \textbf{b} Normalized deflection time $\tau_df$ as a function of $V$. The data collapse on a single curve. The legend is the same as in a.}
\end{figure}

Our setup allows us to measure the deformation of the membrane and to monitor the time $\tau_d$ at which maximal deflection $\delta = \delta_{max}$ occurs. Figure \ref{fig3_SI}a shows the variation of $\tau_d$ as a function of $V$. $\tau_d$ is roughly constant although it slightly increases at small $V$ as observed for $\tau$. The variation of $\tau_d$ with $f_m$ is of more interest, $\tau_d$ decreases as $f_m$ increases. Our coupled oscillator model enables us to plot the rescaled deflection time $\tau_df$ as a function of impact velocity (figure \ref{fig3_SI}b). All the data collapse on a single curve meaning that the timescale of the first oscillation is proportional to that of the full rebound. We find that $\tau$ is roughly equal to twice $\tau_d$.

\section{Membrane modelling}
We model the PDMS sheets as two dimensional ($a \sim$ cm and $h=20$ \micro m)  elastic ($E = 1$ MPa, $\rho = 965$ kg/m$^3$) membranes under tension. 
The membranes are deposited on frames of radius $a$ and put into tension by adding a mass $m$ to the clamps maintaining the edge of the PDMS sheets. Table \ref{table1_SI} summarises the characteristics of the membranes studied. \\
The tension $T$ applied to the membrane is $T = \frac{mg}{2\pi a}$.
The fundamental frequency of a circular membrane under tension $T$ is:
\begin{equation*}
f_m^{th} =\frac{\chi_{01}}{2\pi a}\sqrt{\frac{T}{\mu}}
\end{equation*}
where $\chi_{01}$ is the first 0 of the Bessel function of the first kind and $\mu$ is the area density of the PDMS sheet.
\begin{table}[h]
\centering
\begin{tabularx}{0.7\textwidth}{>{\raggedright}X >{\centering}X >{\centering}X >{\raggedleft\arraybackslash}X}
  \hline
  $a$ (mm) & $m$ (g) & $f_m$ (Hz) & $f_m^{th}$ (Hz) \\
  \hline
  7.5 & 3.41 & 302 & 309 \\
  7.5 & 12.23  & 444 & 586 \\
  10 & 3.41 & 210 & 201 \\
  10 & 12.23 & 290 & 379 \\
  17.5 & 13.5 & 128 & 172 \\
  25 & 4.68 & 60 & 59\\
  25 & 13.5 & 75 & 101 \\
  25 & 29.2  &  96.8  & 148 \\
  \hline
\end{tabularx}
\caption{\label{table1_SI} Characteristics of the eight membranes used in the accompagnying paper: radius $a$, mass $m$, measured and predicted frequencies $f_m$ and $f_m^{th}$ respectively.}
\end{table}

\begin{figure}[h]
\centering
\includegraphics[width = 0.5\textwidth]{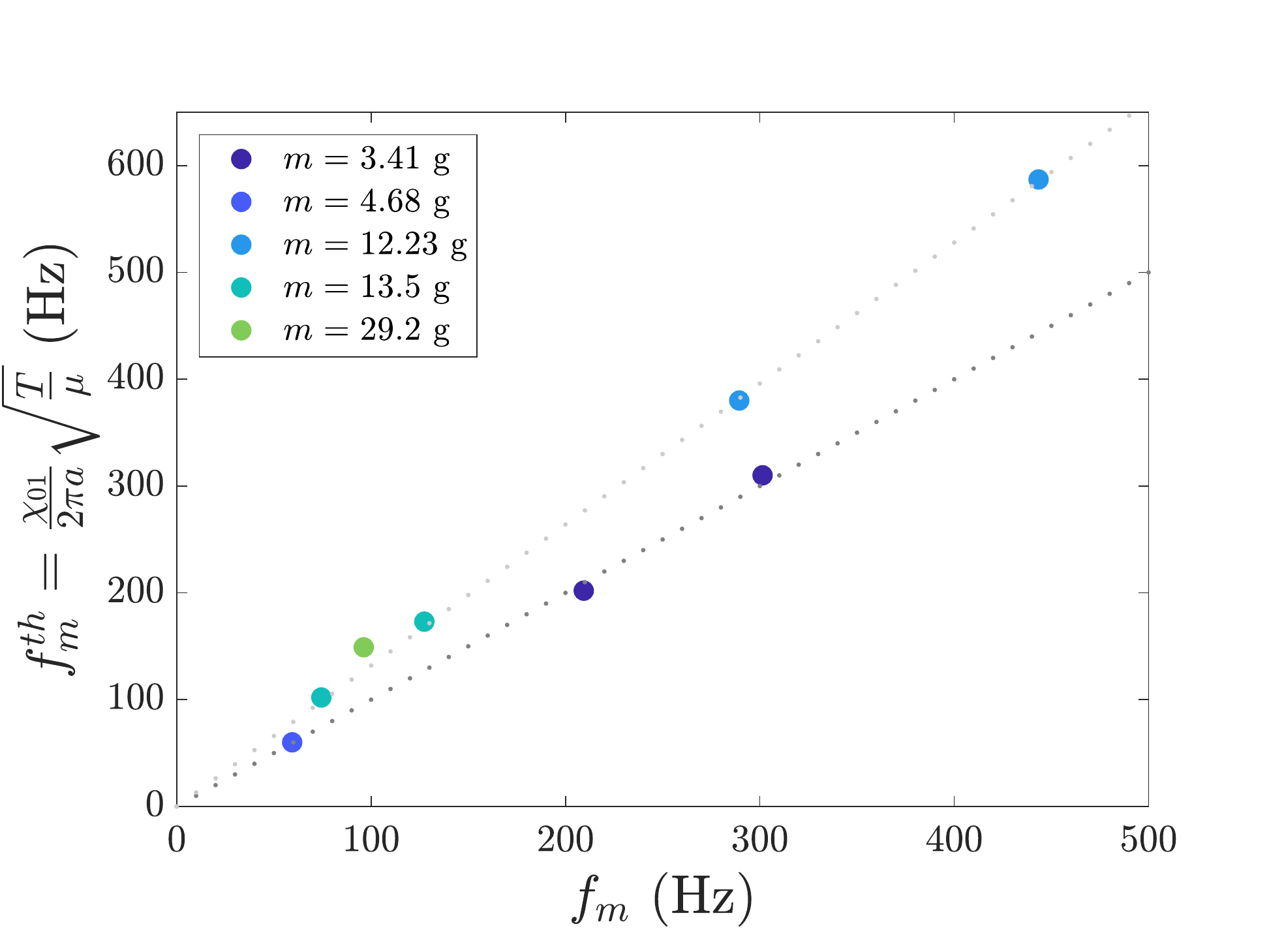}
\caption{\label{fig1_SI}Predicted oscillating frequency $f_m^{th}$ of the membrane compared to the fundamental frequency $f_m$ measured during free oscillations. The grey dotted line has a slope 1 and shows quantitative agreement between measurement and prediction for the low $m$ data. The light grey line has a slope 1.3, it highlights that while we overpredict the frequency a linear relationship between measurement and prediction is kept.}
\end{figure}
Figure \ref{fig1_SI} compares the frequencies measured during free oscillations of the membrane to the above prediction. For low imposed masses $m$ (3.41 and 4.68 g), we have a quantitative agreement between the predicted and measured frequencies. For larger $m$ ($m>3.41$ g), we predict higher frequencies than the measurements but interestingly the linear relationship between measurement and prediction is still observed when the mass is kept constant (light grey dotted line). That overprediction of the oscillating frequency may come from friction between the membrane and the frame, although talc powder is placed on the frame to reduce it, and/or from stretching of the membrane that leads to a reduced effective tension.

 
\section{Contact time of marbles of different $R$}
\begin{figure}[h]
\centering
\includegraphics[width = 0.5\textwidth]{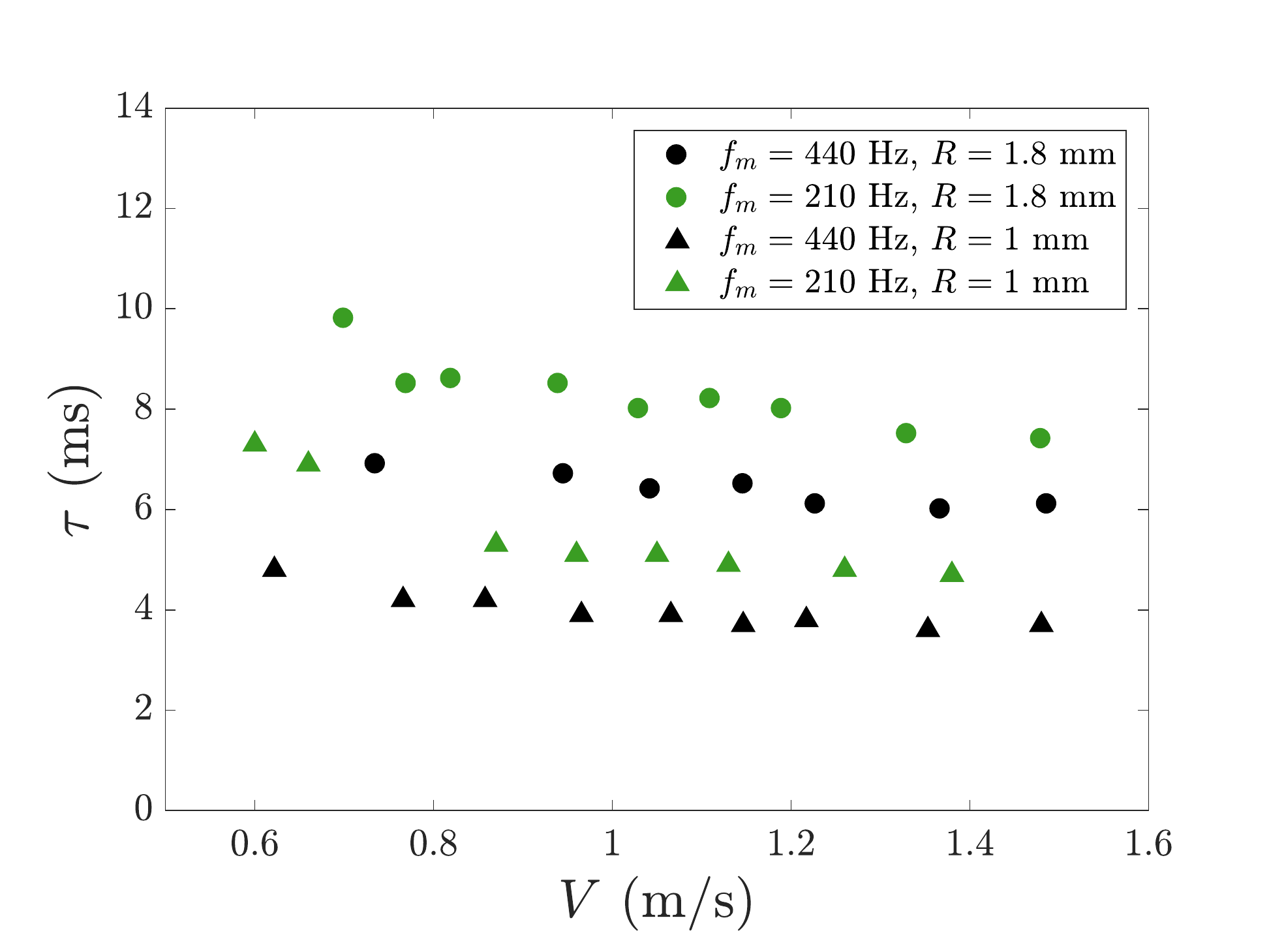}
\caption{\label{fig4_SI}Contact time $\tau $ of marbles as a function of impact velocity $V$. $\tau$ depends on both $f_m$ and $R$.}
\end{figure}

We measure the contact time of marbles with two different radii, $R = 1.8$ and $R = 1$ mm, for two distinct membrane frequencies at which contact time reduction occurs ($\tau_m<\tau_0$). We plot in figure \ref{fig4_SI} the contact time $\tau$ as a function of impact velocity $V$. $\tau$ is roughly constant with $V$ and it decreases with increasing $f_m$ and with decreasing $R$.
For the two frequencies $f_m$, the contact time is increased by a factor $1.3 \pm 0.1$ when varying $R$ by a factor $1.8$. This increase is in good agreement with our scaling analysis that predicts a change of $\tau$ as $R^{3/4}$, that is, a factor $1.55$ when $R$ varies from $1$ to $1.8$ mm, in stark contrast with the expected results from the inertio-capillary scaling that yields a factor $2.4$ ($R^{3/2}$).

\section{Normalized contact time $\tau f_b$ of liquid marbles}
\begin{figure}[h]
\centering
\includegraphics[trim = {0 2.7cm 0 2.7cm},clip,width = 0.7\textwidth]{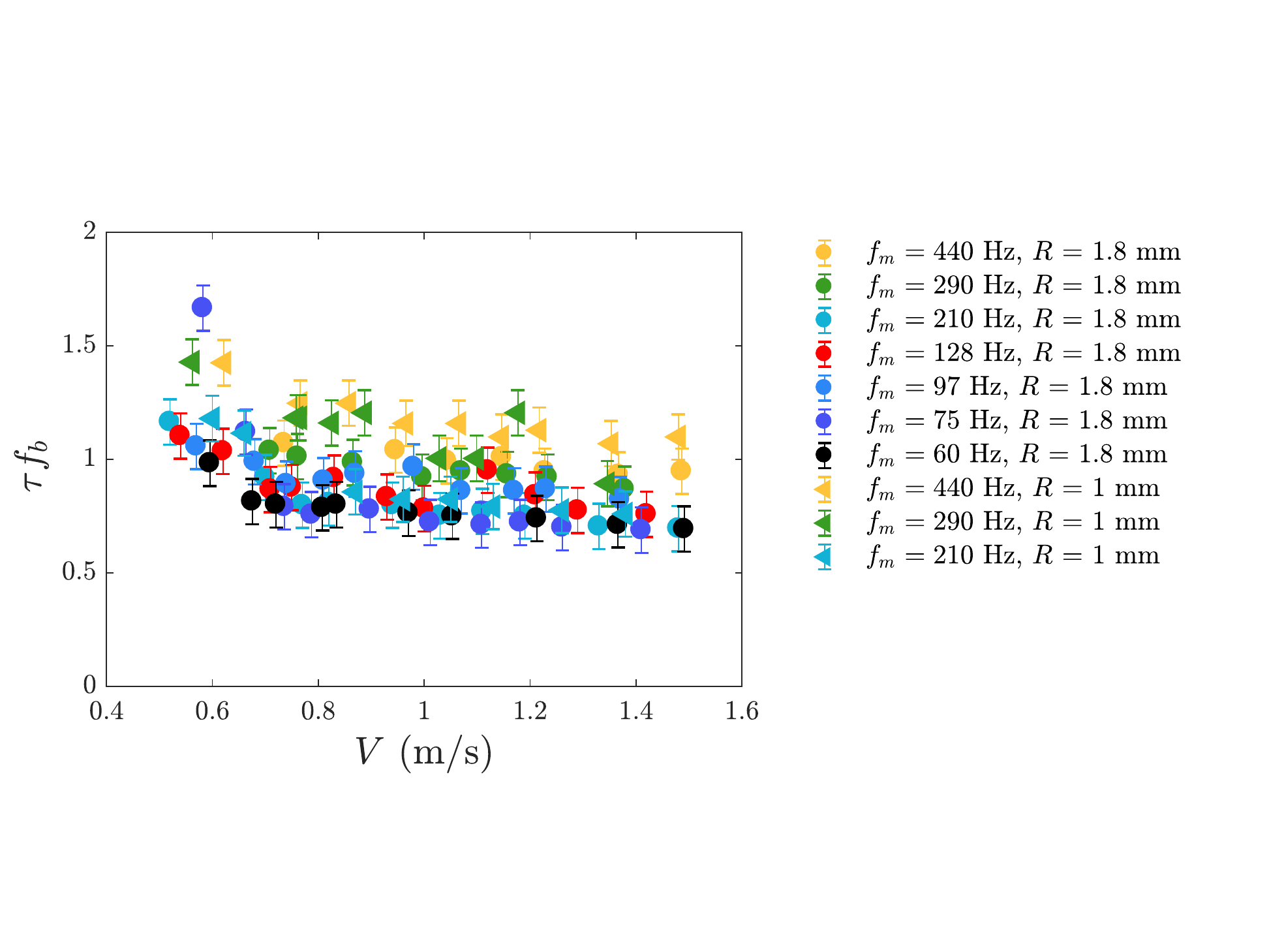}
\caption{\label{fig5_SI}Normalized contact time $\tau f_b $ of liquid marbles impacting a flexible membrane as a function of impact velocity $V$ where $f_b$ is the frequency of a membrane with stiffness $k$ and mass $m_d+m_m$. Data does not collapse suggesting that the deformability of the liquid has to be modelled.}
\end{figure}

The contact time of a rigid bead impacting a flexible membrane is determined by the timescale $1/f_b$, where $f_b = \frac{1}{2\pi}\sqrt{\frac{k}{m_m+m_d}}$ is the frequency of a membrane with stiffness $k$ and mass $m_m + m_d$. Figure \ref{fig5_SI} represents the contact time $\tau$ of drops rescaled by $f_b$ as a function of impact velocity $V$. Data does not collapse on a single curve suggesting that an added mass argument is not enough to model the system and that the deformability of the liquid has to be taken into account.

\section{Natural frequencies of the drop-membrane system}
In our model, the position $z$ of the membrane obeys the equation:
\begin{equation}
\frac{d^4z}{dt^4} + \frac{k_d}{m_m}(1+\frac{m_m}{m_d}+\frac{k}{k_d})\frac{d^2z}{dt^2} + \frac{kk_d}{m_dm_m}z =0
\label{eq1}
\end{equation}
From (\ref{eq1}) we deduce two natural frequencies of the system, that is $f_* = \frac{1}{2\pi}(\frac{k_d}{m_m}(1+\frac{m_m}{m_d}+\frac{k}{k_d}))^{1/2}$ and $f = \frac{1}{2\pi}(\frac{kk_d}{m_dm_m})^{1/4}$.
Let us show that we always have $f_*>f$. Assuming this inequality, we get $\frac{k_d}{m_m}(1+\frac{m_m}{m_d}+\frac{k}{k_d})^2 > \frac{k}{m_d}$. Introducing $\alpha= m_m/m_d$ and $\beta = k/k_d$, this can be rewritten $(1+\alpha+\beta)^2>\alpha\beta$, which indeed is obeyed for positive $\alpha$ and $\beta$.

\section{Movies} 
\noindent
\textbf{Movie S1}: Impact of a liquid marble with radius $R=1.8$ mm on a rigid superhydrophobic susbtrate at velocity $V=0.75$ m/s. We measure $\tau=22.5$ ms. \\
\textbf{Movie S2}: Impact of a liquid marble with radius $R=1.8$ mm on a flexible superhydrophobic substrate ($a=10$ mm, $m=12.23$ g) at velocity $V=1$ m/s. The contact time is $\tau=7$ ms, reduced by 70\% compared to $\tau_0$. The lycopodium grains dispersed in air during impact allow us to visualize the airflow after takeoff.\\
\textbf{Movie S3}: Impact of a rigid bead with radius $R_b = 1.75$ mm on a flexible superhydrophobic substrate ($a=10$ mm, $m=3.41$ g) at velocity $V=1.25$ m/s. We measure $\tau = 6.6$ ms.